\documentstyle[aps,preprint,epsf]{revtex}
\tightenlines
\begin{document}
\draft
\title{Optimized quantum-optical communications\\
in the presence of loss}
\author{G. M. D'Ariano and M. F. Sacchi}
\address{Dipartimento di Fisica \lq Alessandro Volta\rq, Universit\`a 
degli Studi di Pavia,\\
INFM --- Sezione di Pavia\\	
via A. Bassi 6, I-27100 Pavia, Italy\\
{\tt DARIANO@PV.INFN.IT}\hspace{4mm}{\tt MSACCHI@PV.INFN.IT} \\
Phone Number: +39-382-507678\\
Fax Number: +39-382-507563}
\maketitle
\narrowtext
\begin{abstract}
We consider the effect of loss on quantum-optical communication channels.
The channel based on direct detection of number states, which for a
lossless transmission line would achieve the ultimate quantum channel
capacity, is easily degraded by loss. The same holds true for the
channel based on homodyne detection of squeezed states, which also is
very fragile to loss. On the contrary, the ``classical'' channel based
on heterodyne detection of coherent states is loss-invariant. We
optimize the {\em a priori} probability for the squeezed-state and the
number-state channels, taking the effect of loss into account. 
In the low power regime we achieve a sizeable improvement of the
mutual information, and both the squeezed-state and the number-state
channels overcome the capacity of the coherent-state channel. In
particular, the squeezed-state channel beats the classical channel for
total average number of photons $N<8$. However, for sufficiently high power 
the classical channel always performs as the best one.
For the number-state channel we show that with a loss $\eta\lesssim .6$ 
the optimized {\em a priori} probability departs from the usual
thermal-like behavior, and develops gaps of zero probability, with a
considerable improvement of the mutual information 
(up to 70 \% of improvement at low power for attenuation $\eta=.15$). 
\end{abstract}
\section{Introduction}
The detrimental effect of loss is a serious problem for optical
communications based on transmission of nonclassical states of
radiation. 
It is well-known that the results for the lossless case 
\cite{hol,yuoz,caves} rapidly do not 
hold anymore for increasing losses \cite{caves,hirota,hall}. 
As a matter of fact, as shown in this paper, 
the ``nonclassical'' channels based on 
direct detection of number states and homodyning of squeezed 
states---channels that have been originally proposed in order to improve the 
capacity of the ``classical'' channel based on heterodyning of 
coherent states---both are much more sensitive to loss than the 
classical channel. They also have been shown \cite{hall} to be easily 
degraded by additive Gaussian noise, which 
models any kind of environmental effect due to linear interactions 
with random fields. Hence, for long haul communications the
great advantage of using nonclassical states is completely lost, since a
minimum loss of 0.3dB/km is unavoidable with the current
optical-fiber technology. 
In the above scenario the optimization of the quantum channel in the
presence of loss is the most relevant issue for achieving reliable
communication schemes in practical situations. 
\par Through a systematic approach, in this paper we evaluate the 
optimal {\em a priori} probability in the presence of loss, 
for both the squeezed-state and the number-state channels, 
and compare the relative effectiveness in terms of mutual information. 
Although for sufficiently high average 
transmitted power even the optimized channels are anyway beaten by 
the heterodyne one, at low power levels the enhancement of the mutual 
information from optimization makes both nonclassical channels more
effective than the heterodyne one. As we will see in the following,
such improvement of the nonclassical channels is even more dramatic
for very strong attenuation and gives rise to unexpected results. 
\par
\begin{figure}[hbt]\begin{center}
\epsfxsize=.8\textwidth\leavevmode\epsffile{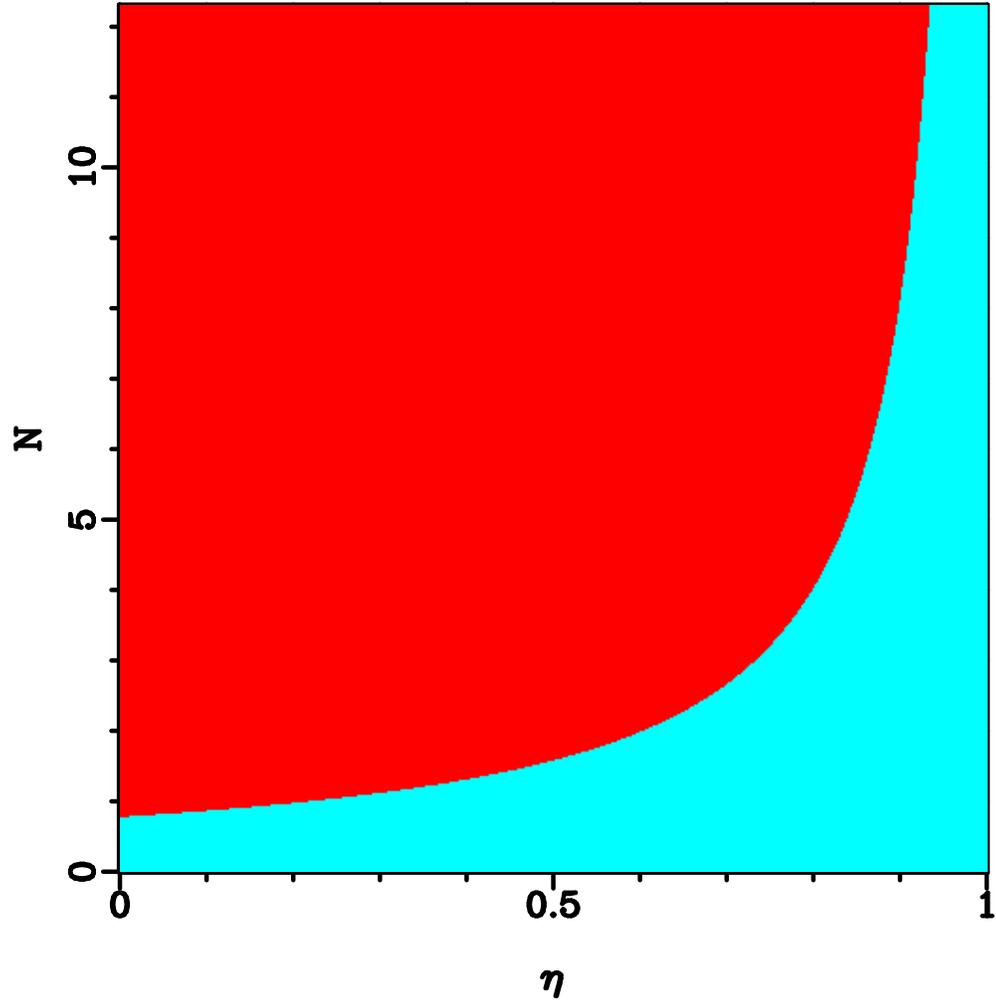}
\end{center}
\caption[fake]{\footnotesize Optimality capacity diagram with 
$\eta$-independent optimization. Black region: the coherent-state 
channel is optimal; dark grey region: 
the number-state channel is optimal.} 
\label{f:comp2}\end{figure}

\begin{figure}[thb]\begin{center}
\epsfxsize=.8\textwidth\leavevmode\epsffile{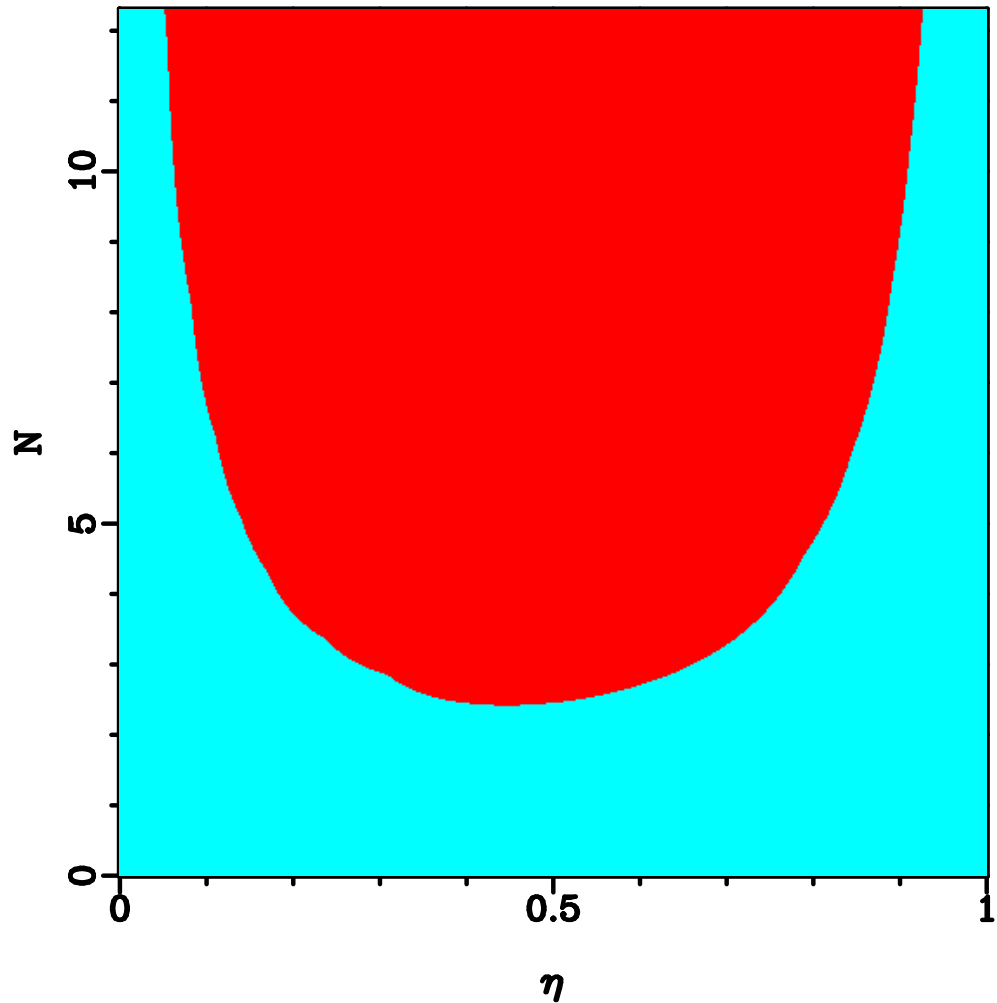}
\end{center}
\caption[fake]{\footnotesize Optimality capacity diagram comparing the 
coherent-state with the optimized number-state channels. In the dark
grey region the optimized number-state channel achieves a superior capacity,
whereas in the black region it is the coherent-state channel the
optimal one.} 
\label{f:comp3}\end{figure}

\begin{figure}[thb]\begin{center}
\epsfxsize=.8\textwidth\leavevmode\epsffile{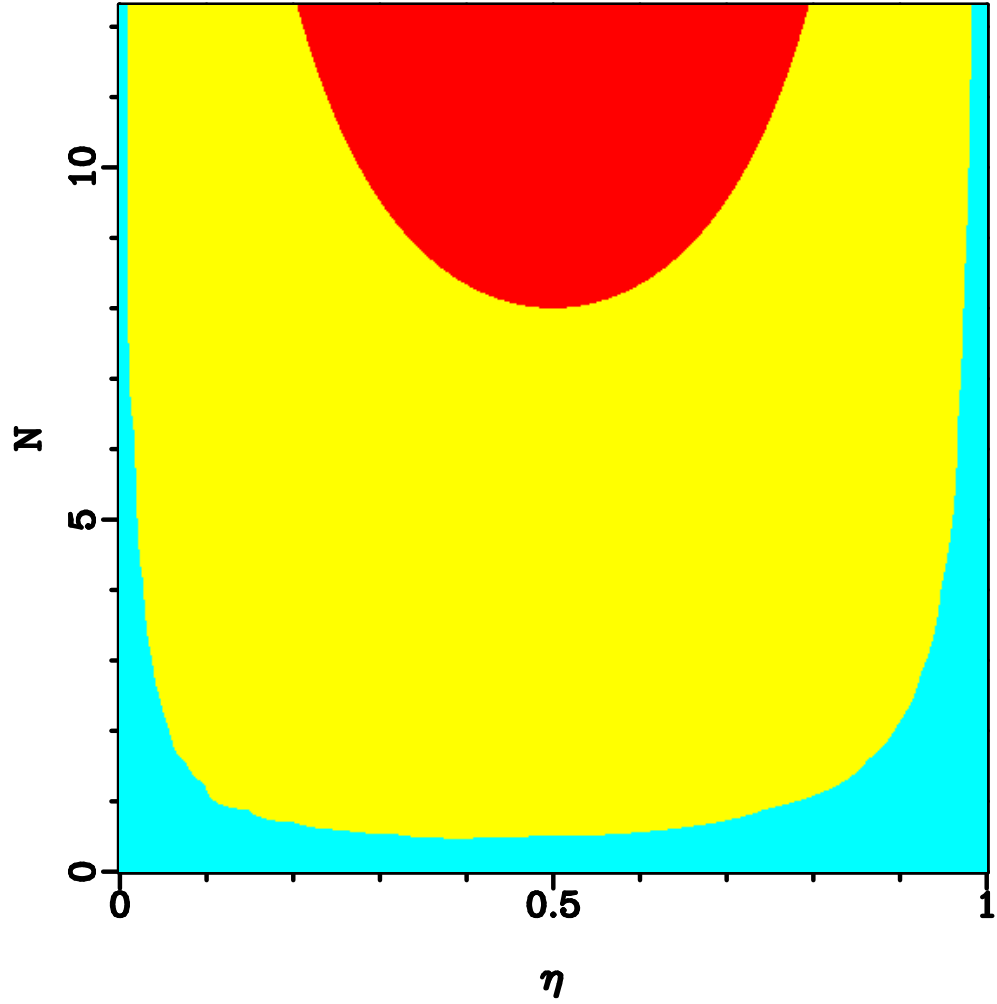}
\end{center}
\caption[fake]{\footnotesize Optimality capacity diagram. In the black
region the coherent-state channel has the highest capacity, 
in the light grey region the best channel is the 
optimized squeezed-state one. Finally,
in the dark grey region the optimal channel is the optimized
number-state one.}  
\label{f:bestopt}\end{figure}

 The paper is organized as follows. In Sect. II we introduce the
master equation that models the effect of loss, and we shortly review 
the heterodyne channel. Due to the peculiar form of the master
equation---which keeps coherent states as coherent---in the presence of
loss there is no need for optimization of the {\em a priori}
probability, whereas the channel capacity only depends on the average
photon number at the receiver. In Sect. III the optimal {\em a priori}
probability for the squeezed-state channel is derived analytically. We
show that the optimal fraction of squeezing photons rapidly decreases
with loss, with a relative improvement of the mutual information
up to 30 \% at low power for $\eta=.15$. For total mean photon number
$N < 8$ the optimized squeezed-state channel beats the coherent-state
one at any value of the loss. Following Hall \cite{exc}, we also 
provide a general upper bound valid for any lossy channel that uses
homodyne detection, a bound that, however, is never achieved by our
optimized squeezed-state channel. Sect. IV is devoted to the
optimization of the number-state channel. Using the recursive Blahut's
algorithm \cite{blahut}, we obtain an optimized {\em a priori} probability 
that departs from the usual monotonically-decreasing thermal-like
behavior, and that, for attenuation $\eta\lesssim .6$,
develops gaps of zero probability at intermediate numbers of photons. 
An intuitive explanation of this result can be understood as the effect 
of a loss so strong that it becomes more convenient to use 
a smaller alphabet of well-spaced letters in order to achieve a 
better distinguishability at the receiver. 
The sizeable improvement of the mutual information---over 70 \% for 
high attenuation at low power---partially stems the detrimental
effect of loss. In Sect. V the main conclusions are drawn. The paper
is accompanied by many optimality capacity diagrams
(Figs. \ref{f:bestcs}, \ref{f:bestcso}, \ref{f:comp2}, \ref{f:comp3}  
and \ref{f:bestopt}), which compare the different communication channels,
giving the regions in the loss-power plane where each channel is optimal
with respect to the others.
\section{Heterodyne channel}
The communication channel based on heterodyne detection encodes 
a complex variable on a coherent state with Gaussian {\em a priori}
distribution. The heterodyne 3dB detection noise is itself Gaussian
additive, and the Gaussian form of the {\em a priori} probability density 
that achieves the channel capacity is dictated by the Shannon's
theorem\cite{caves,gall} for Gaussian channels subjected to the
quadratic constraint of fixed average power. Under such constraint the
variance of the optimal Gaussian distribution equals the value of the
mean photon number $N$. In the following we briefly redraw the
analytical derivation of this result, in order to show how the optimal
{\em a priori} probability remains unchanged in the presence of loss.
\par The effect of loss on a single-mode communication channel is
determined by the master equation
\begin{eqnarray}
\partial _t\hat\varrho={\cal L}_{\Gamma}\hat\varrho 
\doteq\Gamma(n_a+1)L[a]\hat\varrho +
\Gamma n_a L[a^{\dag}]\hat\varrho \simeq \Gamma L[a]\hat\varrho 
\;,\label{loss}\end{eqnarray}   
where the superoperator ${\cal L}_{\Gamma}$ gives the time derivative of 
the density matrix $\hat\varrho$ of the radiation state (in the 
interaction picture) through the action of the Lindblad superoperators
$L[a]\hat\varrho=a\hat\varrho a^{\dag}-{1\over 2} 
(a^{\dag}a\hat\varrho+\hat\varrho a^{\dag}a)$ \cite{lind}. 
The coefficient $\Gamma$ represents the damping rate, whereas 
$n_a$ denotes the mean number of thermal photons at the 
frequency of mode $a$, and can be neglected 
at optical frequencies. We introduce the energy attenuation factor, or
``loss'', defined as follows
\begin{eqnarray}
\eta\doteq\exp(-\Gamma t)\;,\label{etag}
\end{eqnarray}
according to the evolution of the average power
\begin{eqnarray}
\langle a^{\dag} a(t)\rangle \equiv \hbox{Tr}[a^{\dag} a\,\hat\varrho (t)]
=\hbox{Tr}[a^{\dag} a
\,e^{{\cal L}_{\Gamma}t}\hat\varrho (0)]=\eta\langle a^{\dag} a (0)\rangle \;. 
\end{eqnarray}
More generally, $\eta $ gives the scaling factor of any normal-ordered 
operator function, namely
\begin{eqnarray}
e^{{\cal L}^{\vee }_{\Gamma}t}\mbox{\bf :} f(a^{\dag} ,a)
\mbox{\bf :} =\mbox{\bf :} f(\eta ^{1/2}a^{\dag} ,\eta
^{1/2}a)\mbox{\bf :} \;,\label{ord}
\end{eqnarray}
where ${\cal L}^{\vee }_{\Gamma}$ denotes the dual Liouvillian, which
is defined through the identity
\begin{eqnarray}
\mbox{Tr} \left[ (e^{{\cal L}^{\vee }_{\Gamma}t}\hat O)\hat\varrho \right]
=\mbox{Tr}\left[\hat O (e^{{\cal L}_{\Gamma}t}\hat\varrho )\right]  \; 
\end{eqnarray}
valid for any operator $\hat O$. The mutual information transmitted
throughout the channel for {\em a priori} distribution $p(\alpha )$
of the encoded complex variable $\alpha $, and for input-output conditional
probability density $Q(\beta |\alpha )$, is given by 
\cite{hol,yuoz,caves,gall}
\begin{eqnarray}
I=\int d^2 \alpha\, p(\alpha ) \int d^2 \beta\, Q(\beta|\alpha)
\ln{Q(\beta|\alpha)\over 
\int d^2 \alpha '\,p(\alpha ')Q(\beta|\alpha ')}\;,\label{infoc} 
\end{eqnarray}
where the integrations are performed on the complex plane with 
measure $d^2\alpha =d\, \hbox{Re}\alpha \,d\, \hbox{Im}\alpha $. 
For heterodyning of a coherent state $|\alpha \rangle $, 
the conditional probability density is given by
\begin{eqnarray}
Q(\beta |\alpha )=|\langle \beta |\alpha \rangle |^2={1\over \pi}\exp
\left( -|\beta -\alpha |^2\right) \;.\label{psc}
\end{eqnarray}
In the presence of loss $\eta$, according to Eqs. (\ref{loss}) and
(\ref{etag}) one has  
$e^{{\cal L}_{\Gamma}t}(|\alpha\rangle\langle\alpha |)=
|\eta ^{1/2}\alpha\rangle\langle\eta ^{1/2}\alpha |$, 
and hence the conditional probability density simply rewrites
\begin{eqnarray}
Q_\eta (\beta |\alpha )={1\over \pi}\exp
\left( -|\beta -\eta ^{1/2}\alpha|^2\right) \;.\label{psc2}
\end{eqnarray}
The constraint of fixed average power at the transmitter reads
\begin{eqnarray}
\int d^2 \alpha\,p(\alpha )\langle \alpha |a^{\dag} a|\alpha \rangle  = 
\int d^2 \alpha\,p(\alpha )\,|\alpha |^2=N\;,\label{cons}
\end{eqnarray}
where in the following $N$ will generally denote the total mean
photon number. We now maximize the mutual information
(\ref{infoc}) over all possible normalized probability densities
$p(\alpha )$ that satisfy the constraint
(\ref{cons}). Eq. (\ref{infoc}) can be simplified as follows 
\begin{eqnarray}
I=-\ln \pi -1 -\int d^2\beta \,f(\beta )\ln f(\beta )
\;,\label{infos}
\end{eqnarray}
where $f(\beta )$ denotes the unconditioned or ``{\em a posteriori}'' 
probability, namely 
\begin{eqnarray}
f(\beta )=\int d^2 \alpha \,p(\alpha )\,Q_\eta (\beta |\alpha )
\;.\label{conv}
\end{eqnarray}
By transferring the (normalization and power) constraints from
$p(\alpha )$ to $f(\beta )$, we can maximize the mutual information
with respect to $f(\beta )$ through a variational calculus on
Eq. (\ref{infos}). While normalization condition for $p(\alpha )$
simply corresponds to normalization of $f(\beta )$, the fixed-power
constraint needs the following algebra
\begin{eqnarray}
N&=&\int d^2\alpha \,p(\alpha )\,|\alpha |^2
= \int {d^2\beta\over \eta}\int d^2\alpha\,p(\alpha )\,|\alpha |^2
{1\over \pi}\exp\left(-{|\beta |^2\over \eta }-
\eta |\alpha |^2 +\beta \bar\alpha 
+\bar\beta \alpha \right)\nonumber \\
&=&{1\over \eta}\int d^2\beta\,(|\beta |^2-1) f(\beta )\;, 
\end{eqnarray}
where the bar denotes the complex conjugate number. 
Hence, the variational equation for the mutual information writes
\begin{eqnarray}
0={\delta\over \delta f}\left[ I-\lambda 
\int d^2\beta \,f(\beta )-{\mu\over \eta}
\int d^2\beta \,(|\beta |^2-1)f(\beta )\right] 
\;,\label{var}
\end{eqnarray}
with $I$ given by Eq. (\ref{infos}), and with $\lambda $ and $\mu $
as Lagrange multipliers to be determined. It is easy found that
Eq. (\ref{var}) has the Gaussian solution 
\begin{eqnarray}
f(\beta )={1\over \pi(\eta N +1)}\exp\left( -{|\beta |^2\over \eta N +1}
\right ) \;,\label{sol}
\end{eqnarray}
and from Eqs. (\ref{infos}) and (\ref{sol}) one obtains the capacity of the 
heterodyne channel in the presence of loss
\begin{eqnarray}
C=\ln (1+\eta N)\;. 
\end{eqnarray}
Hence, the channel capacity depends only on the mean photon number
$\eta N$ at the receiver. Eqs. (\ref{conv}) and (\ref{sol}) give the
optimal {\em a priori} probability density 
\begin{eqnarray}
p(\alpha )={1\over \pi N}\exp\left( -{|\alpha |^2\over N}\right) 
\;,\label{palfa}
\end{eqnarray}
which is manifestly independent on $\eta$, with the consequence that the
optimal {\em a priori} probability for the lossless heterodyne channel is
still optimal in the presence of loss. This result is due 
to the peculiar form of the master equation (\ref{loss}), which keeps
coherent states as coherent. As we will show in the following, this
will no longer hold true for the squeezed-state and the number-state
channels.
\section{Homodyne channel}
\begin{figure}[htb]
\begin{center}
\epsfxsize=.8\textwidth\leavevmode\epsffile{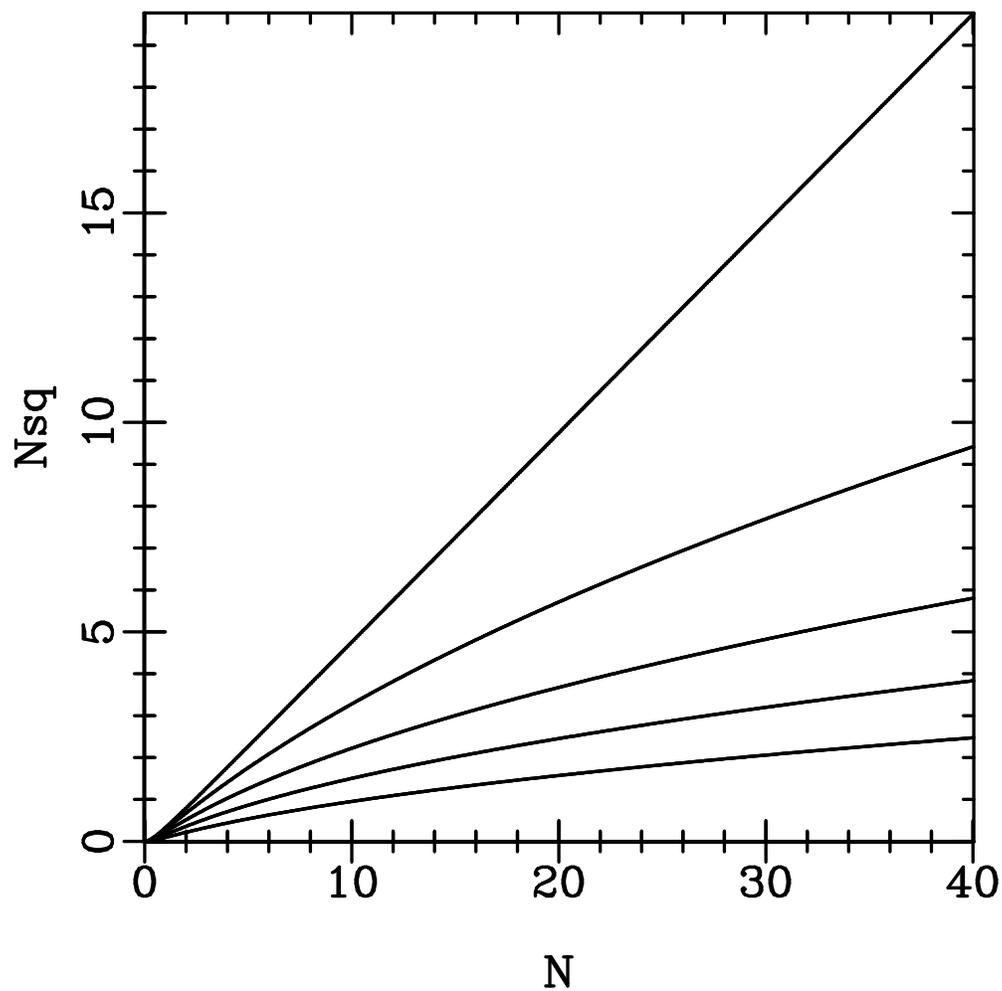}
\end{center}
\caption[fake]{\footnotesize 
Number of squeezing photons that optimizes the lossy homodyne channel 
versus the total average number of photons, at different values of 
the attenuation factor $\eta $. From the top to the bottom, 
the plotted lines refer to $\eta =1,\ .95,\ .85,\ .7,\ .5$.}
\label{f:npsq}\end{figure}

\begin{figure}[htb]\begin{center}
\epsfxsize=.8\textwidth\leavevmode\epsffile{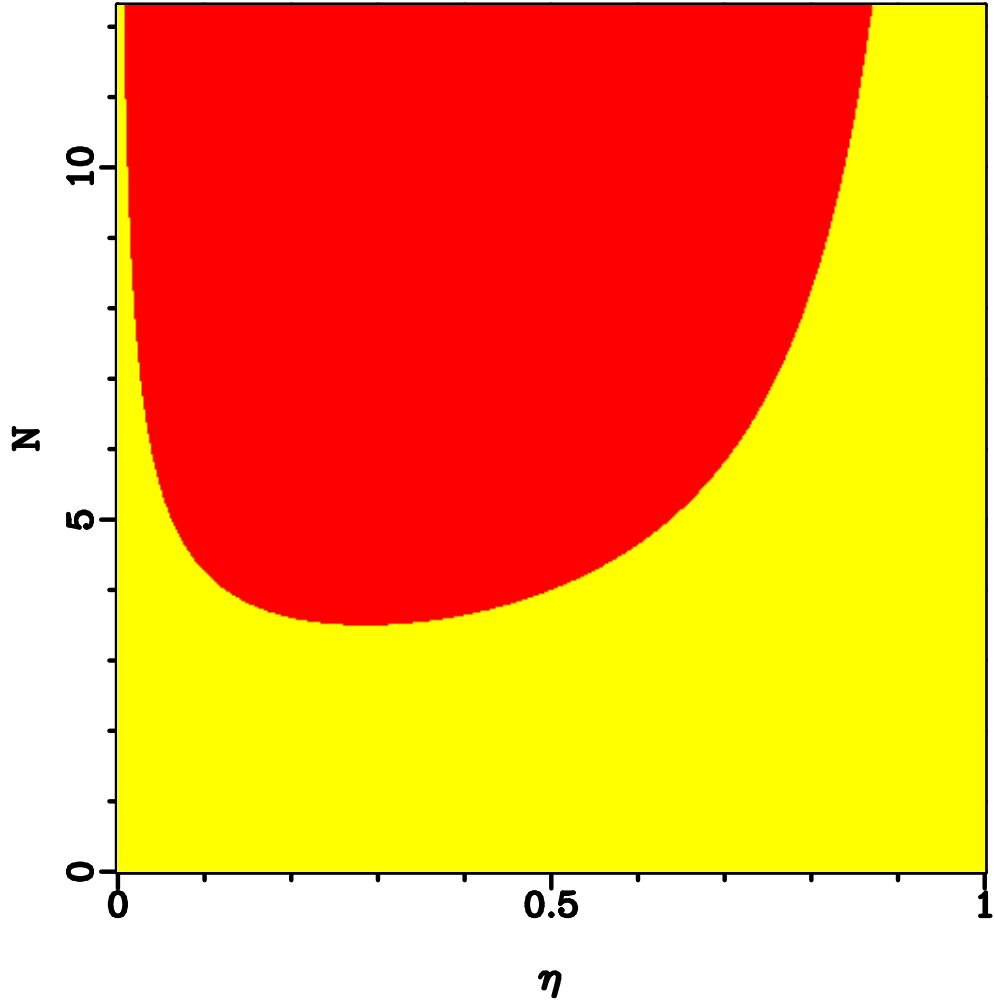}
\end{center}
\caption[fake]{\footnotesize Optimality capacity diagram, which
represents the region where the coherent-state channel is optimal
(black area) and that where the squeezed-state channel is optimal
instead (green area). Both channel are the customary ones, which were
optimized for the lossless case.}
\label{f:bestcs}\end{figure}

\begin{figure}[thb]\begin{center}
\epsfxsize=.8\textwidth\leavevmode\epsffile{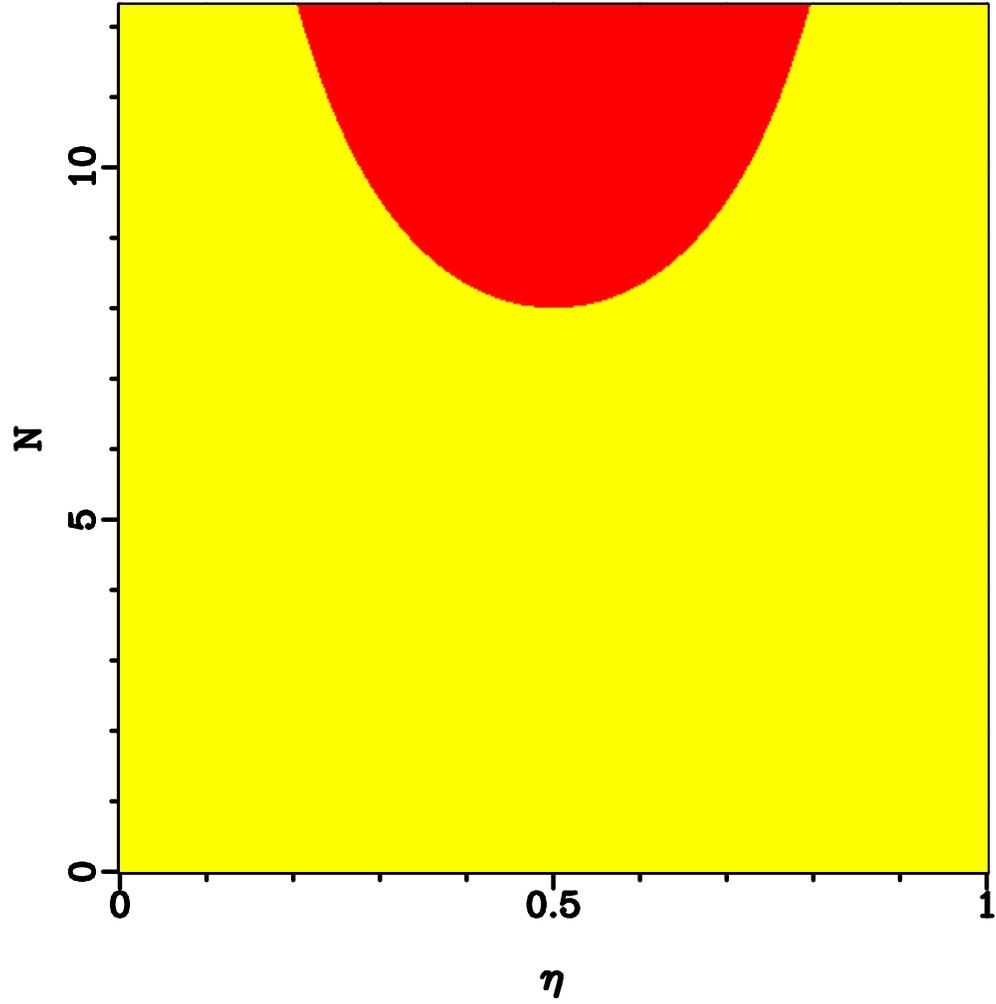}
\end{center}
\caption[fake]{\footnotesize Optimality capacity diagram comparing the
coherent-state channel to the squeezed-state channel in the presence
of loss. Among the two channels, in the grey region the squeezed state
channel has the highest capacity, whereas in the black region the
coherent state channel is the best.}
\label{f:bestcso}\end{figure}

\begin{figure}[htb]
\begin{center}
\epsfxsize=.8\textwidth\leavevmode\epsffile{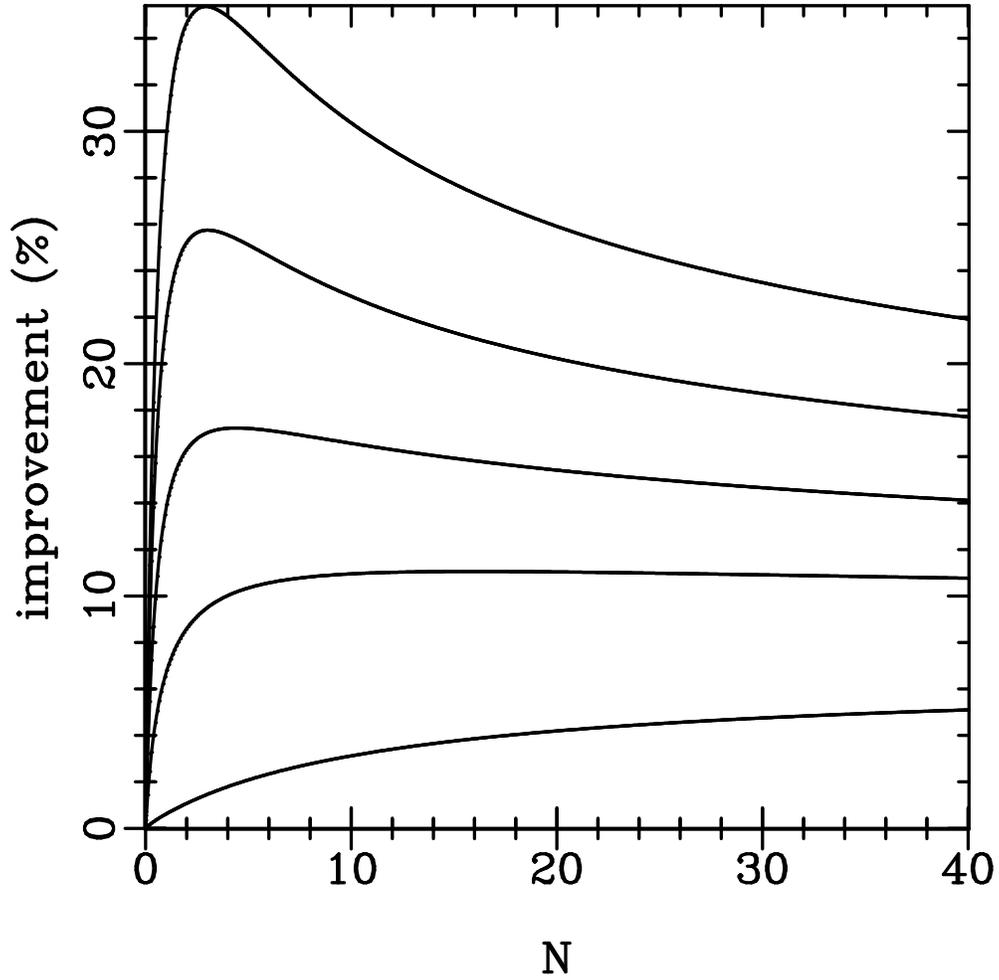}
\end{center}
\caption[fake]{\footnotesize {\em Per cent} improvement of the mutual 
information versus the total average number of photons $N$, 
with $\eta$-independent optimization. 
The plotted lines refer to different values of the attenuation 
factor $\eta $. From the top to the bottom $\eta =.15,\ .25,\ .4,\
.6,\ .9$.} 
\label{f:impsq}\end{figure}

The homodyne channel encodes a real variable $x$ on the 
quadrature-squeezed state
\begin{eqnarray}
|x\rangle _{r}=D(x)S(r)|0\rangle \;,\label{qs}
\end{eqnarray}
which is generated from the vacuum $|0\rangle $ through the action of
the displacement operator $D(x)$ and of the squeezed operator $S(r)$,
which are defined as follows
\begin{eqnarray}
&&D(x)=\exp\left[ x\left( a^{\dag} -a\right) \right] \\
&&S(r)=\exp\left[{r\over 2}\left( {a^{\dag} }^2-a^2\right)\right] \;. 
\end{eqnarray}
The decoding is performed by homodyning a fixed quadrature, say 
$\hat X\equiv (a+a^{\dag})/2$. For lossless transmission, the conditional 
probability density of getting the value $x'$ when the transmitted state 
is $|x\rangle _r$ writes
\begin{eqnarray}
Q(x'|x)=|\langle x'|x\rangle _r|^2=\sqrt{{1\over 2\pi\Delta ^2}}\exp\left[ 
-{(x'-x)^2\over 2\Delta^2}\right] \;, 
\end{eqnarray}
where $|x'\rangle $ denotes the eigenstate of $\hat X$, and the 
variance is given by $\Delta ^2=e^{-2r}/4$. 
According to Shannon's theorem\cite{caves,gall} the optimal {\em a
priori} probability $p(x)$ for the ideal homodyne channel has the
Gaussian form
\begin{eqnarray}
p(x)=\sqrt{{1\over 2\pi\sigma ^2}}\exp\left(-{x^2\over 2\sigma ^2} 
\right) \;,\label{pix}
\end{eqnarray}
with variance
\begin{eqnarray}
\sigma ^2={N(N+1)\over 2N+1}\;.\label{sigma}
\end{eqnarray}
The fixed-power constraint is given by
\begin{eqnarray}
N=\int dx\, p(x){}_r\langle x|a^{\dag} a|x\rangle _{r}
=\int dx\,p(x)(x^2+\sinh ^2 r)=\sigma ^2 +\sinh ^2 r\;.\label{fix}
\end{eqnarray}
Hence, Eq. (\ref{sigma}) corresponds to fix the fraction of squeezing 
photons at the value $\sinh ^2 r=N^2/(2N+1)$.
The capacity is given by
\begin{eqnarray}
C=\int dx\int dx'\,p(x)Q(x'|x)\ln{Q(x'|x)\over 
\int d\tilde x\, p(\tilde x)Q(x'|\tilde x)}
={1\over 2}\ln \left(1+{\sigma ^2\over \Delta ^2} \right)=\ln (1+2N) 
\;. 
\end{eqnarray}
In the presence of loss, by means of the identity (\ref{ord}) and the 
following normal-ordered representation of the quadrature projector
\begin{eqnarray}
|x\rangle \langle x|=\delta (\hat X -x)=\int {d \lambda \over 2\pi}
\,e^{-i\lambda x}\,e^{-\lambda ^2/8}\,e^{i\lambda a^{\dag} /2}\,e^{i\lambda a/2}
\;,\label{proj}
\end{eqnarray}
one obtains the conditional probability density
\begin{eqnarray}
&&Q_{\eta }(x'|x)=\langle x'|e^{{{\cal L}_{\Gamma}t}}
(|x\rangle _{r}{}_{r}\langle x|)|x'\rangle =\mbox{Tr} \left[e^{{\cal L}^{\vee }_{\Gamma}t}
\left(\delta (\hat X -x')\right)|x\rangle _{r}{}_{r}\langle x| \right] \nonumber \\
&&=\mbox{Tr} \left[ \int {d\lambda \over 2\pi}\,e^{-i\lambda x'}\,
e^{-\lambda ^2(1-\eta )/8}\,e^{i\eta ^{1/2}\lambda \hat X}
|x\rangle _{r}{}_{r}\langle x|\right] =
\sqrt{{1\over 2\pi\Delta _{\eta }^2}}\exp\left[ 
-{(x'-\eta ^{1/2} x)^2\over 2\Delta _{\eta }^2}\right]
\;,\label{alg}
\end{eqnarray}
where 
\begin{eqnarray}
\Delta _{\eta }^2={1\over 4}\left[1-\eta\left(1-e^{-2r}\right)\right]
\;. 
\end{eqnarray}
For Gaussian {\em a priori} probability with variance $(N-\sinh ^2 r)$, which
satisfies the fixed-power constraint (\ref{fix}), the mutual
information is given by  
\begin{eqnarray}
I={1\over 2}\ln \left( 1+{4\eta (N-\sinh ^2 r)\over 1-\eta 
(1-e^{-2r})}\right) \;.\label{infsq}
\end{eqnarray}
Upon maximizing Eq. (\ref{infsq}) with respect to $\xi \equiv e^{-2r}$
we obtain
\begin{eqnarray}
I={1\over 2}\ln \left( 1+{4\xi N-(1-\xi)^2\over \xi ^2 +{1- \eta\over
\eta}\xi}\right) \; 
\end{eqnarray}
with 
\begin{eqnarray}
\xi ={\eta +\sqrt{1+4\eta (1-\eta )N}\over (4N +1)\eta +1}\;. 
\end{eqnarray}
The optimal number of squeezing photons is given by $(\xi +1/\xi
-2)/4$, and it is plotted versus $N$ in Fig. \ref{f:npsq}
for some values of the attenuation $\eta $. One can see that
the optimal fraction of squeezing photons rapidly decreases with
attenuation. 
This means that for increasing loss it is more and more unprofitable 
to use much power to squeeze the quadrature of the signal, since 
the quantum noise of the state at the receiver approaches to that of the 
coherent state. These results agree with previous investigations on  
the loss effects in terms of signal-to-noise ratio\cite{hirota}.  
Figs. \ref{f:bestcs} and \ref{f:bestcso} are optimality capacity diagrams, 
which compare different channels giving the regions on the loss-power
plane where each channel is optimal. The coherent-state channel is
compared to the squeezed-state channel without and with loss-dependent
optimization in Fig. \ref{f:bestcs} and Fig. \ref{f:bestcso},
respectively. One can see that the optimization leads to a sizeable
improvement of the mutual information, especially for strong
attenuation and low power (see also Fig. \ref{f:impsq}), making
the diagram symmetric around the $\eta=1/2$ vertical axis. Notice the
location of the minimum at $\eta =.5$ and $N=8$ on the boundary
between the optimality regions in Fig. \ref{f:bestcso}: this means
that for mean power less than eight photons the squeezed-state
channel always beats the coherent-state one, independently on
attenuation. 
\par Through a kind of exclusion principle for the information contents of
quantum observables \cite{exc}, Hall has proved an upper bound for the
information that can be achieved by a homodyne channel subjected to
Gaussian noise. Following Hall's method, here we prove the following
upper bound for any lossy channel that uses homodyne detection
\begin{eqnarray}
I\leq \ln (1+2\eta N)\;.\label{hbound}
\end{eqnarray}
By denoting with $S(\hat A|\hat\varrho)$ the entropy associated to the 
probability distribution $\langle a|\hat\varrho|a\rangle$ of the
eigenvalue $a$ of the observable $\hat A$ when the state is
$\hat\varrho$, namely 
\begin{eqnarray}
S(\hat A|\hat\varrho)=-\int da\, \langle a|\hat\varrho|a\rangle\,\ln
\langle a|\hat\varrho|a\rangle\;, 
\end{eqnarray}
the mutual information retrieved from the measurement of the
observable $\hat A$ on a member of the ensemble specified by the
density matrix $\hat\varrho =\sum _i p_i \hat\varrho _i$ is given by
\begin{eqnarray}
I(\hat A|\hat\varrho)=S(\hat A|\hat\varrho)-\sum _i p_i 
S(\hat A|\hat\varrho _i)\;.\label{in}
\end{eqnarray}
A simple variational calculation gives the upper bounds
\begin{eqnarray}
&&S(\hat X|\hat\varrho)\leq {1\over 2}+{1\over 2}\ln (2\pi \langle 
\Delta\hat X^2\rangle _{\hat\varrho}) 
\;\label{up}\\
&&S(\hat Y|\hat\varrho)\leq {1\over 2}+{1\over 2}\ln (2\pi \langle 
\Delta\hat Y^2\rangle _{\hat\varrho})
\;\label{1up}
\end{eqnarray}
for the entropy associated to the conjugated quadratures 
$\hat X=(a+a^{\dag} )/2$ and $\hat Y=(a-a^{\dag} )/2i$, the notation
$\langle\ldots\rangle_{\hat\varrho}$ representing the ensemble average 
with density operator $\hat\varrho$. 
\par\noindent Moreover, writing $\hat\varrho $ as a mixture of pure states
\begin{eqnarray}
\hat\varrho =\sum _j p_j|\psi _j\rangle \langle \psi _j|\;, 
\end{eqnarray}
from the concavity of entropy one has
\begin{eqnarray}
S\left(\hat X|e^{{\cal L}_{\Gamma}t}\hat\varrho \right )=
S\left(e^{{\cal L}^{\vee }_{\Gamma}t}\hat X|\hat\varrho \right )
\geq \sum _j p_j S\left( e^{{\cal L}^{\vee }_{\Gamma}t}\hat X|
(|\psi _j\rangle\langle\psi _j |)\right )\geq 
\inf _j S\left( e^{{\cal L}^{\vee }_{\Gamma}t}\hat X|
(|\psi _j\rangle\langle\psi _j |)\right )
\;,\label{2up}
\end{eqnarray}
and analogously for the other quadrature $\hat Y$. A derivation
similar to that of Eq. (\ref{alg}) leads to the conditional
probability
\begin{eqnarray}
&&p\left(x|e^{{\cal L}_{\Gamma}t}  
(|\psi _j\rangle\langle\psi _j |)
\right )=
\mbox{Tr} \left\{ \left[{2\over\pi (1-\eta )} \right]^{1/2}\exp 
\left[ -{2(\eta ^{1/2}\hat X -x)^2\over 1-\eta }\right]    
|\psi _j\rangle\langle\psi _j |
\right\} \nonumber \\
&\equiv& \mbox{Tr} \left[ G\left(\eta ^{1/2}\hat X -x\right) 
|\psi _j\rangle\langle\psi _j |\right]
\;, 
\end{eqnarray}
where we introduced the Gaussian operator-valued measure
defined as follows
\begin{eqnarray}
G(\eta ^{1/2}\hat X -x)\equiv 
\left[{2\over\pi (1-\eta )} \right]^{1/2}\exp 
\left[ -{2(\eta ^{1/2}\hat X -x)^2\over 1-\eta }\right]    
\;. 
\end{eqnarray}
By varying over the bra $\langle \psi _j|$ the following quantity
\begin{eqnarray}
J=S\left( e^{{\cal L}^{\vee }_{\Gamma}t}\hat X|
(|\psi _j\rangle\langle\psi _j |)
\right )+S\left( e^{{\cal L}^{\vee }_{\Gamma}t}\hat Y|
(|\psi _j\rangle\langle\psi _j |)
\right )-\lambda \left(\langle \psi _j|\psi _j\rangle -1\right)
\;, 
\end{eqnarray}
one obtains the variational equation
\begin{eqnarray}
&&0={\delta J\over \delta \langle \psi _j|}=
-\int dx\,G(\eta ^{1/2}\hat X -x)\ln
\left\{ \mbox{Tr}\left[ G(\eta ^{1/2}\hat X -x)
|\psi _j\rangle\langle\psi _j |
\right] \right\} 
|\psi _j\rangle \nonumber\\&&
-\int dy\,G(\eta ^{1/2}\hat Y -y)\ln
\left\{ \mbox{Tr}\left[ G(\eta ^{1/2}\hat Y -y)
|\psi _j\rangle\langle\psi _j |
\right] \right\} 
|\psi _j\rangle -(\lambda +2)|\psi _j\rangle 
\;,\label{gei}
\end{eqnarray}
where $\lambda $ is the Lagrange multiplier for the normalization
constraint relative to the state $|\psi _j\rangle $. It can be easily
verified that the case of vacuum state $|\psi_j\rangle\equiv|0\rangle$
satisfies Eq. (\ref{gei}). Then, from Eq. (\ref{2up}) along with the following
relation
\begin{eqnarray}
S\left( e^{{\cal L}^{\vee }_{\Gamma}t}\hat O|
(|0\rangle\langle 0|)
\right )={1\over 2}
-{1\over 2}\ln \left( {2\over \pi}\right) \; 
\end{eqnarray}
that holds for any quadrature operator $\hat O$, one has
\begin{eqnarray}
S\left(\hat X|e^{{\cal L}_{\Gamma}t}\hat\varrho \right )+
S\left(\hat Y|e^{{\cal L}_{\Gamma}t}\hat\varrho \right )
\geq 1+\ln \left({\pi\over 2}\right)
\;.\label{sotto}
\end{eqnarray}
On the other hand, from Eqs. (\ref{up}) and (\ref{1up}) one obtains
\begin{eqnarray}
S\left(\hat X|e^{{\cal L}_{\Gamma}t}\hat\varrho \right )+
S\left(\hat Y|e^{{\cal L}_{\Gamma}t}\hat\varrho \right )
\leq 1+\ln (2\pi)+{1\over 2}\ln\left( \langle \Delta\hat X^2\rangle _
{e^{{\cal L}_{\Gamma}t}\hat\varrho }
\langle \Delta\hat Y^2\rangle _{e^{{\cal L}_{\Gamma}t}\hat\varrho }\right) 
\;.\label{step}
\end{eqnarray}
The product of the expectation values in Eq. (\ref{step}) 
can be maximized as follows
\begin{eqnarray}
&&\langle \Delta\hat X^2\rangle _
{e^{{\cal L}_{\Gamma}t}\hat\varrho }
\langle \Delta\hat Y^2\rangle _{e^{{\cal L}_{\Gamma}t}\hat\varrho }=
\left[ \eta \langle \Delta\hat X^2\rangle _{\hat\varrho }
+{1\over 4}(1-\eta)\right]
\left[ \eta \langle \Delta\hat Y^2\rangle _{\hat\varrho }
+{1\over 4}(1-\eta)\right]
\nonumber \\
&& \leq {1\over 4}\left[ \eta\left( \langle\hat X^2\rangle 
_{\hat\varrho }+\langle\hat Y^2\rangle _{\hat\varrho }\right) 
+{1-\eta\over 2}\right]^2 \leq 
{1\over 4}\left( \eta \langle a^{\dag} a\rangle _{\hat\varrho}
+{1\over 2}\right)^2 
\;,\label{step2}
\end{eqnarray}
and we obtain
\begin{eqnarray}
S\left(\hat X|e^{{\cal L}_{\Gamma}t}\hat\varrho \right )+
S\left(\hat Y|e^{{\cal L}_{\Gamma}t}\hat\varrho \right )
\leq 1+\ln \left({\pi\over 2}\right)+
\ln\left(1+ 2\eta \langle a^{\dag} a\rangle _{\hat\varrho}\right) \;
.\label{end}
\end{eqnarray}
Finally, inequalities (\ref{sotto}) and (\ref{end}), together with
Eq. (\ref{in}) yield the information exclusion relation
\begin{eqnarray}
I\left(\hat X|e^{{\cal L}_{\Gamma}t}\hat\varrho \right )+
I\left(\hat Y|e^{{\cal L}_{\Gamma}t}\hat\varrho \right )\leq \ln 
\left(1+ 2\eta N \right) 
\;\label{2exc}
\end{eqnarray}
where $N=\langle a^{\dag} a\rangle _{\hat\varrho}$. From
Eq. (\ref{2exc}) the bound (\ref{hbound}) follows as a particular
case. From the above derivation we see that the bound
(\ref{hbound}) holds for any lossy channel that employs homodyne
detection.
\par The upper bound (\ref{hbound}) is trivially achieved for $\eta
=1$ by a Gaussian ensemble of squeezed states, however, in the presence of
loss it is not reached by our optimized channel. As a matter of fact, 
there is still room for a slight improvement of the mutual information
if one allows the squeezing $r$ to vary as a function of the signal $x$
in Eq. (\ref{qs}). However, such further optimization is not
achievable analytically---due to the now non Gaussian form of the
conditional probability density---nor it can be worked out
numerically, as no viable method is at hand. 
\section{Direct-detected channel}
\begin{figure}[htb]
\begin{center}
\epsfxsize=.8\textwidth\leavevmode\epsffile{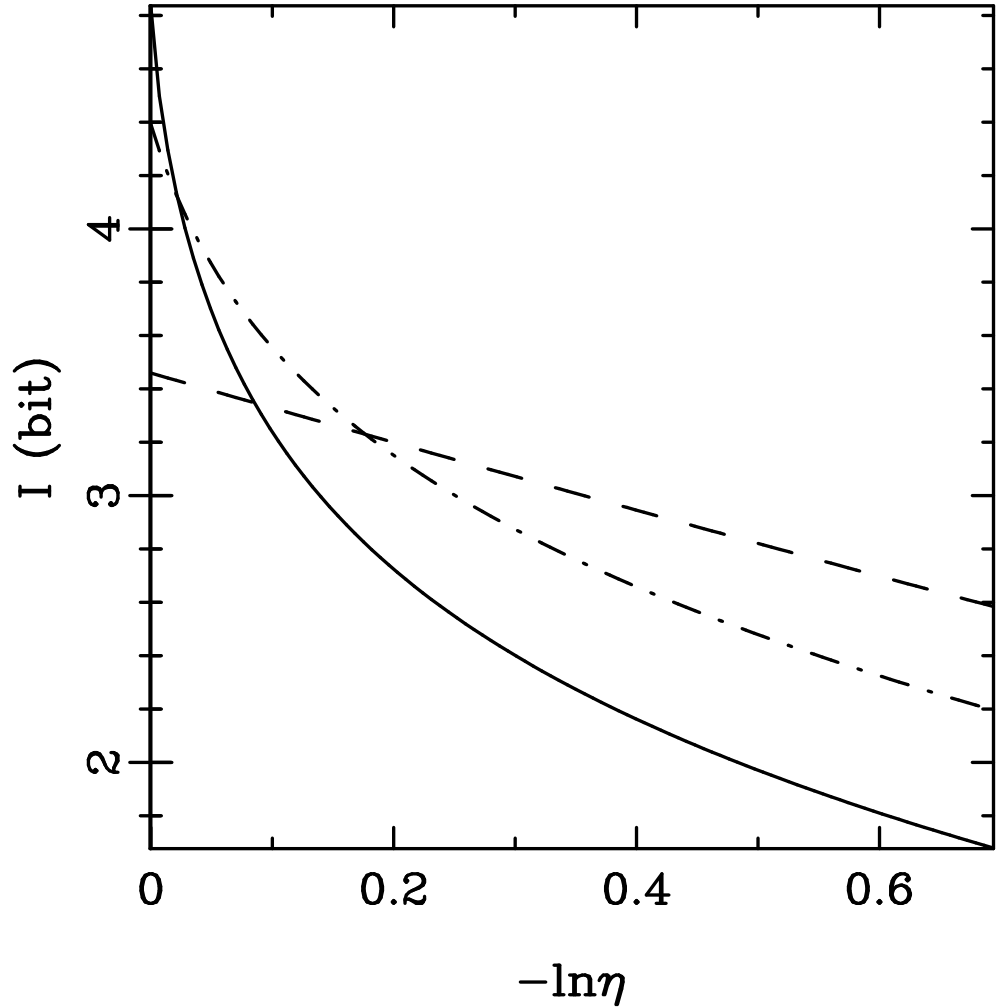}
\end{center}
\caption[fake]{\footnotesize 
Mutual information versus attenuation for the number-state (full), 
the coherent-state (dashed), and the squeezed-state (dashed-dotted) channels. 
The fixed average number of photons is $N=10$. The {\em a priori} 
probability densities are the customary ones for the lossless case
[Eqs. (\ref{pn}), (\ref{palfa}) and (\ref{pix}), respectively].}
\label{f:comp1}\end{figure}

The ideal communication channel that uses direct detection of
Fock-states with thermal {\em a priori} probability
\begin{eqnarray}
p^{\hbox{\scriptsize th}}_n=
{1\over 1+N}\left( {N\over 1+N}\right)^n \;\label{pn}
\end{eqnarray}
achieves the ultimate quantum capacity (the Holevo's bound
\cite{hol,yuoz}) with the constraint of fixed average number of
photons $N$. The ultimate quantum capacity is given by
\begin{eqnarray}
C=\ln\left( 1+N\right)+N\ln\left( 1+{1\over N}\right)  \;. 
\end{eqnarray}
For ideal transmission the conditional probability density is given by 
the Kronecker delta $\delta _{m,n}$. In the presence of loss this is
replaced by the binomial distribution
\begin{eqnarray}
Q_{m,n}(\eta )={n\choose m}\eta ^m\left( 1-\eta\right)^{n-m} \;,\label{bin}
\end{eqnarray}
which represents the probability of detecting $m$ photons when the 
transmitted state is $|n\rangle$. The number-state channel is more
sensitive to loss than the coherent-state and the squeezed-state
ones. In Fig. \ref{f:comp1} the mutual information for the three
channels is plotted versus $\eta $, at fixed power $N=10$, and with
the customary {\em a priori} probabilities optimized for the lossless case
[Eqs. (\ref{palfa}), (\ref{pix}) and (\ref{pn})]. One can see that at
this power level a signal attenuation of 0.5 dB is sufficient to
degrade the number-state channel below the capacity of the
coherent-state channel, whereas at higher power levels the effect is even more
dramatic.  The optimality capacity diagram in Fig. \ref{f:comp2} compares 
the number-state with the coherent-state channels. One can see that in
the presence of loss the number-state channel rapidly loses off its 
efficiency, especially  for high power and strong attenuation.
\par Now we address the problem of optimizing the {\em a priori}
probability distribution in the presence of loss. In principle one
could perform the optimization analytically by varying the information
over the infinite set of variables $\lbrace p_n\rbrace$, however with
no viable method for constraining each $p_n$ to be nonnegative. For
this reason we decided to carry out the optimization numerically,
using the recursive Blahut's algorithm\cite{blahut}. The recursion is
given by
\begin{eqnarray}
c^{(r+1)}_n&=&\exp\left(\sum_k
Q_{k,n}(\eta )\ln\frac{Q_{k,n}(\eta )}
{\sum_m p^{(r)}_m Q_{k,m}(\eta )}-\mu n\right)\;,\nonumber\\
p^{(r+1)}_n&=&p^{(r)}_n\frac{c^{(r)}_n}{\sum_m p^{(r)}_m c^{(r)}_m} 
\;,\label{iter}
\end{eqnarray}
where $p^{(r)}_n$ is the {\em a priori} probability at the $r$th
iteration, $Q_{k,n}(\eta )$ is the conditional probability (\ref{bin}), 
and $\mu$ is the Lagrange multiplier for the average-power
constraint. The series are actually truncated to a finite
dimension, corresponding to a maximum allowed number of photons. 
Blahut proved that the quantity
\begin{eqnarray}
J^{(r)}=I^{(r)}-\mu N^{(r)}\;\label{iter2}
\end{eqnarray}
is increasing versus $r$, and achieves the desired bound, $I^{(r)}$ and
$N^{(r)}$ denoting the mutual information and the average photon
number with the $r$th iterated {\em a priori} probability $p^{(r)}_n$.
For a given $\mu$ one evaluates the limit of $p^{(r)}_n$ for
$r\to\infty$ under the recursion (\ref{iter}), and determines the mutual
information $I$ and the mean photon number $N$ for such limiting
$p^{(\infty)}_n$: in this way the capacity versus power $I=I(N)$ is
obtained as parameterized by $\mu$. 
\par
\begin{figure}[htb]
\begin{center}
\epsfxsize=.85\textwidth\leavevmode\epsffile{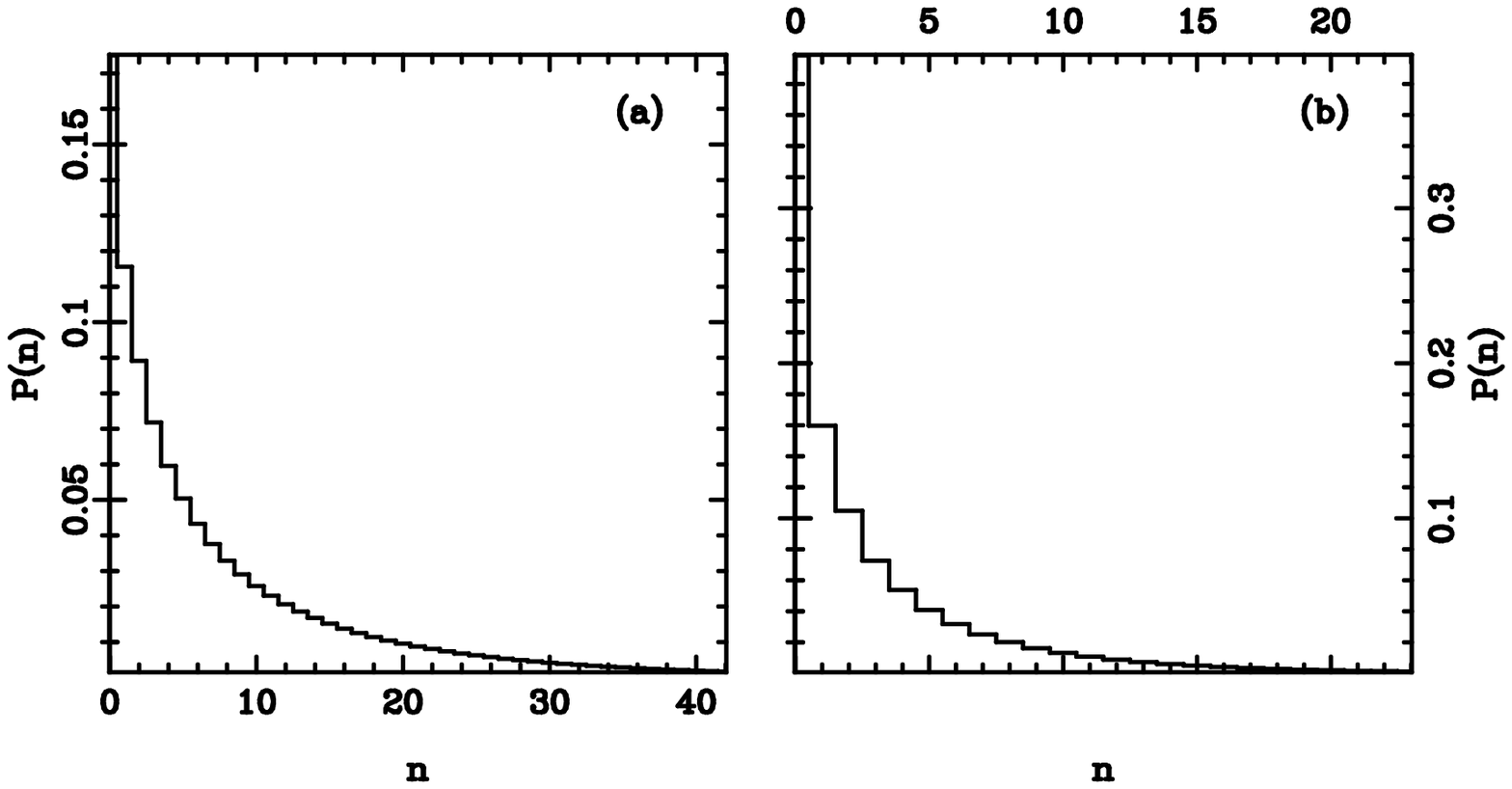}
\end{center}
\begin{center}
\epsfxsize=.85\textwidth\leavevmode\epsffile{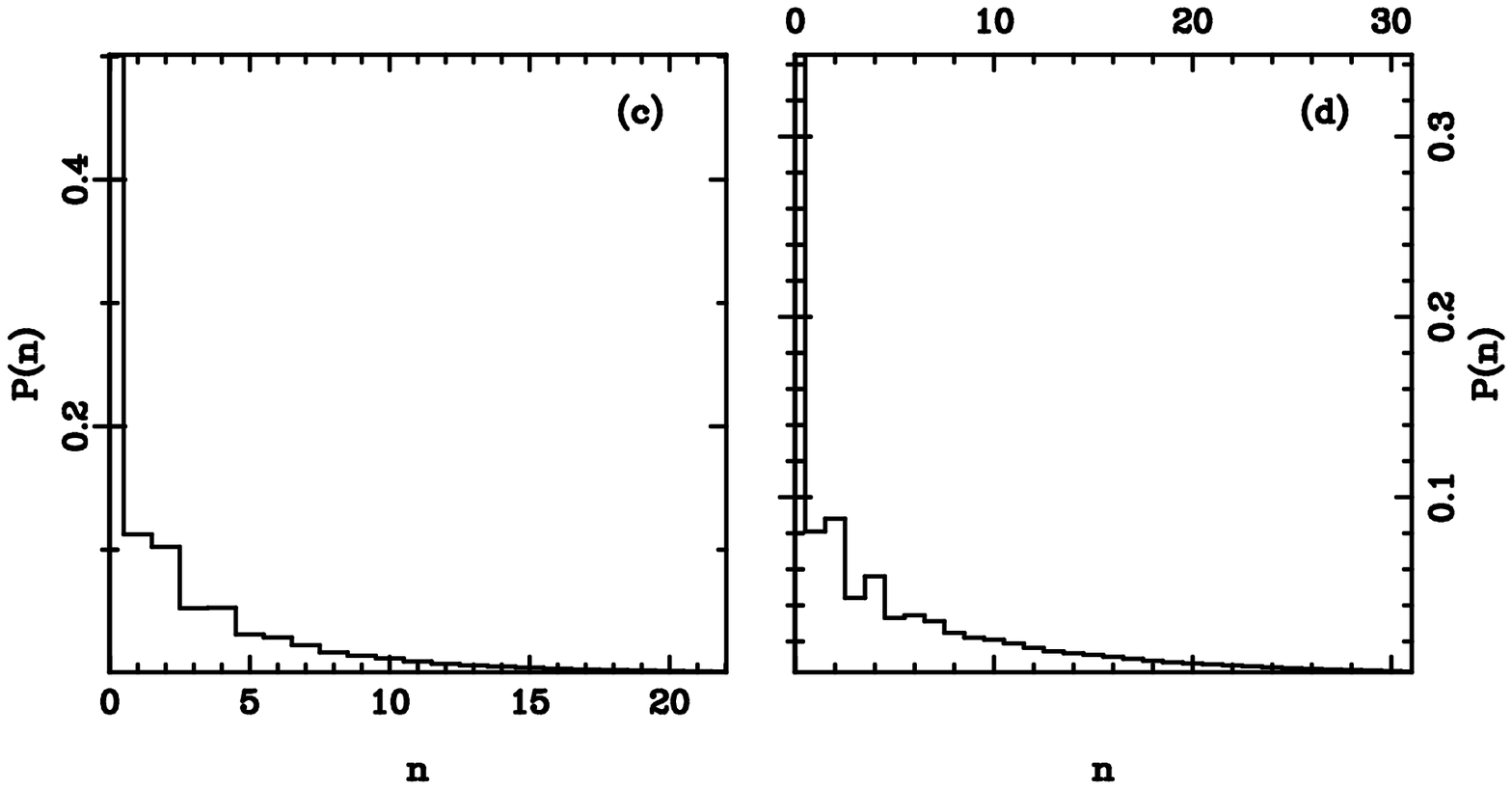}
\end{center}
\begin{center}
\epsfxsize=.85\textwidth\leavevmode\epsffile{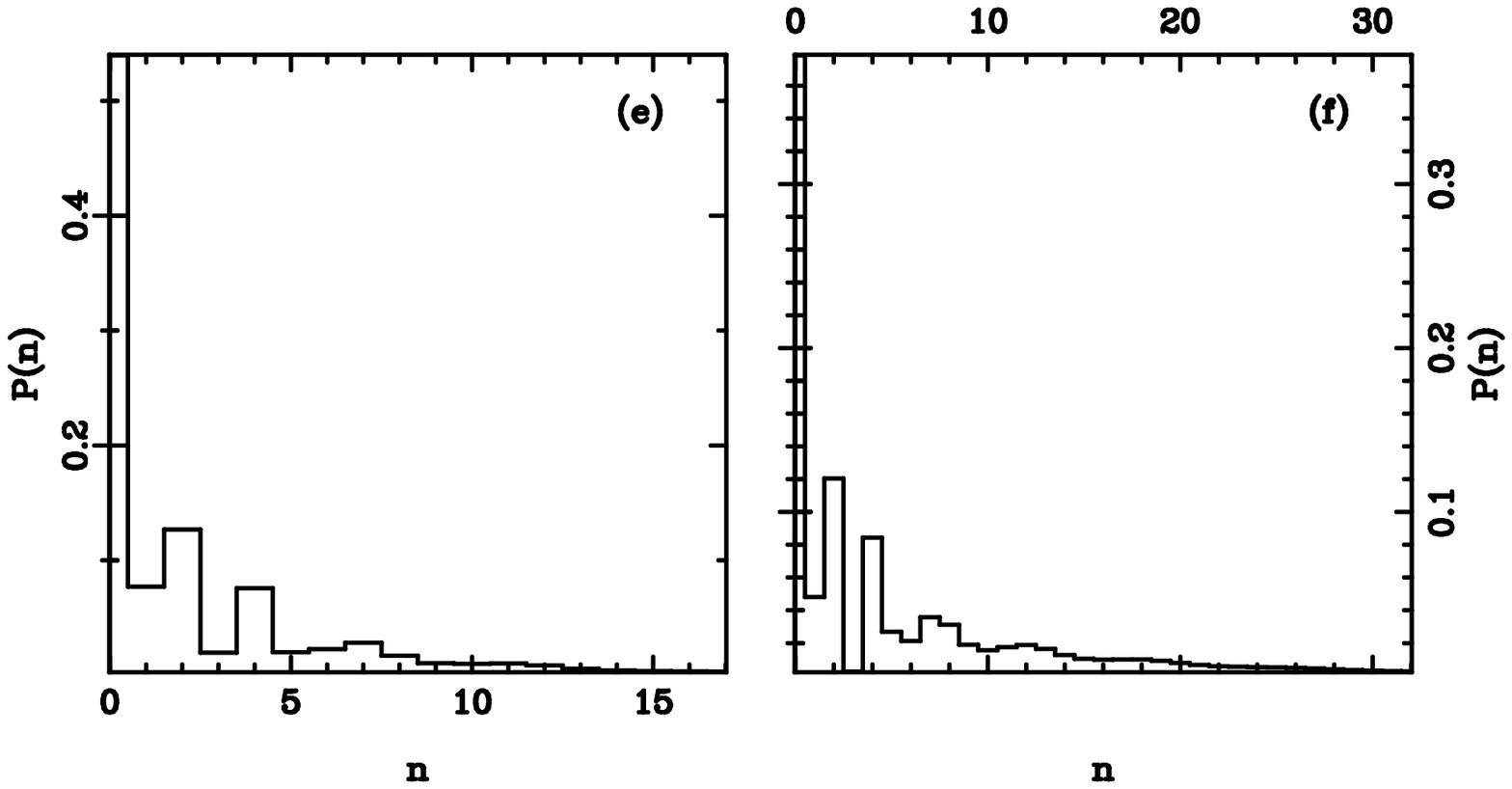}
\end{center}
\begin{center}
\epsfxsize=.85\textwidth\leavevmode\epsffile{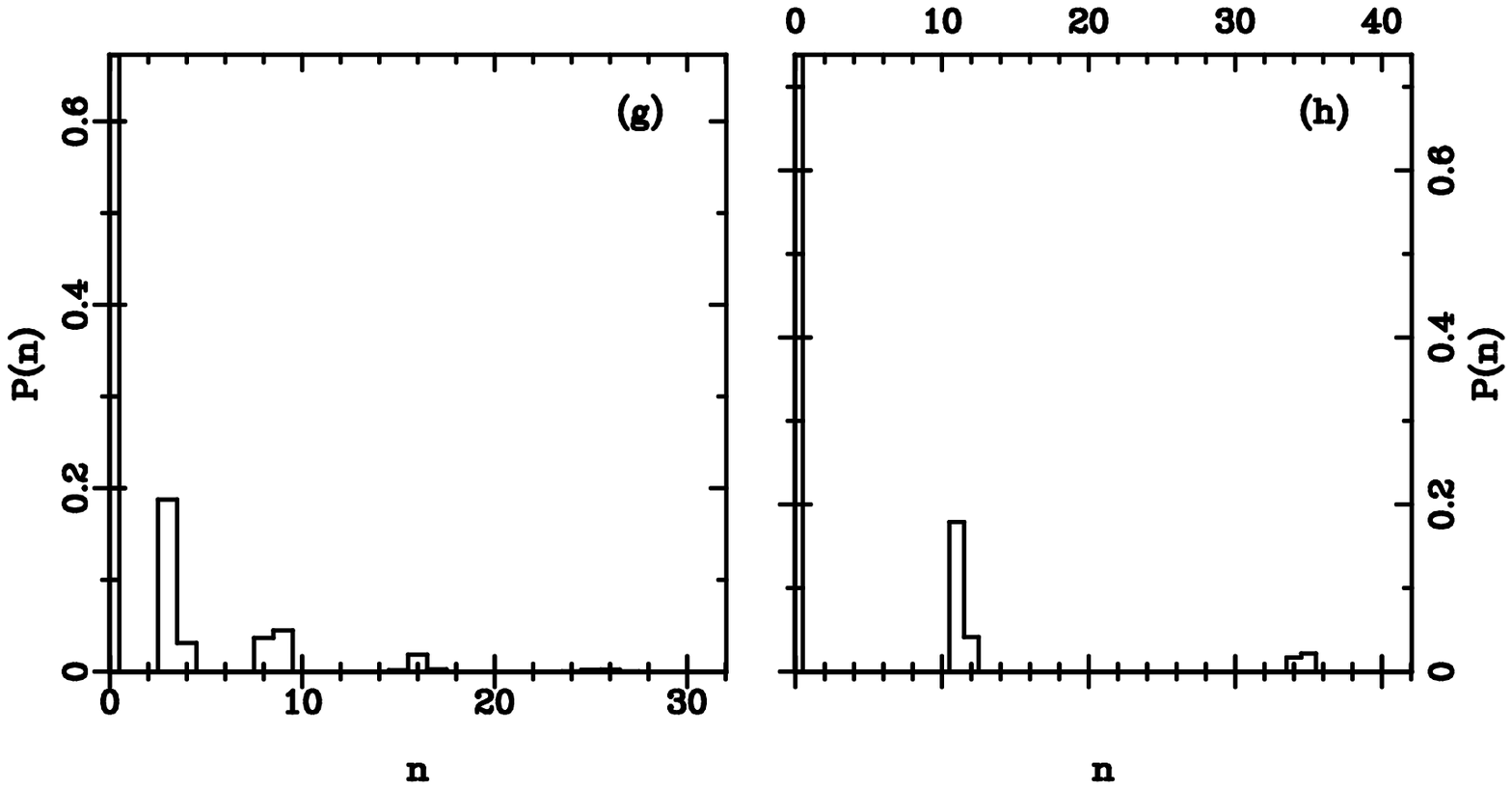}
\end{center}
\caption[fake]{\footnotesize A {\em a priori} probability 
$p(n)$ versus $n$ for different values of 
the attenuation factor $\eta$ and the average power $N$, optimized in
the presence of loss. 
a) $\eta=.9,\ N=8.575$, b) $\eta=.75,\ N=2.872$, 
c) $\eta=.6,\ N=2.414$, d) $\eta=.6,\ N=6.930$ 
e) $\eta=.55,\ N=2.288$, f) $\eta=.55,\ N=6.729$, 
g) $\eta=.4,\ N=1.888$, h) $\eta=.15,\ N=4.040$
[see Table \ref{tab}].}
\label{f:pn1}\end{figure}

 Now we present some numerical results. Figs. \ref{f:pn1} show the
number probability distribution for different 
values of loss and power, evaluated  by means of the Blahut's recursive 
algorithm, stopped at $10^5$ iterations. The Hilbert space has been
truncated at dimension 200, however, truncation at 100 gives almost
identical results. For stronger loss, the optimal {\em a priori}
probability departs from the thermal-like behavior, with an enhanced
vacuum probability. For loss $\eta \lesssim .6$ (see Fig. \ref{f:pn1})
the probability plot develops gaps of zero probability at intermediate
numbers of photons. This can be intuitively understood as the effect
of a loss so strong that it becomes more convenient to use 
a smaller alphabet of well-spaced letters in order to achieve a
better distinguishability at the receiver.
The increase of the probability pertinent the vacuum state comes 
clearly from the constant-energy constraint. Table \ref{tab} provides a
list of numerical results pertaining Figs. (\ref{f:pn1}). It gives the
{\em per cent} improvement of the mutual information after
optimization, along with the absolute value of the mutual information for 
the optimized number-state channel, for the number-state channel with 
customary thermal probability and for the coherent-state channel 
at given value of the loss and of the mean photon number. 
Also the values of the quantities $\epsilon_I$ 
and $\epsilon_P$ are reported, for convergence estimation  of the 
Blahut's recursion (\ref{iter2}). They are defined as the increment 
$\epsilon_I=J^{(r)}-J^{(r-1)}$ of the quantity $J^{(r)}$ in 
Eq. (\ref{iter2}), and the distance $\epsilon _P=\mbox{max}_n 
|p^{(r)}_n-p^{(r-1)}_n|$ between probability plots, both $\epsilon_I$ 
and $\epsilon_P$ being evaluated at the last iteration step 
$r=10^5$. One can see that, according to the small values of $\epsilon
_I$ and $\epsilon _P$, the algorithm is converging quite fast 
(indeed only 10 steps are usually sufficient to get an 
estimate of the capacity up to the second digit). With the occurrence 
of gaps in the {\em a priori} probability, the relative improvement of 
the mutual information increases even more dramatically, up to $70 \%$ 
for strong attenuation $\eta=.15$. At low power, this improvement allows 
the direct-detection channel to overcome the coherent-state channel 
capacity [see Figs. \ref{f:pn1}a,c,e,g,h and their pertaining numerical 
values in Table \ref{tab}]. The optimality capacity diagram in 
Fig. \ref{f:comp3} compares the optimized number-state channel with
the coherent-state channel. Notice the difference with respect to
Fig. \ref{f:bestcso}: here the optimized number-state channel beats the
heterodyne channel at power much lower than for the optimized squeezed-state
channel in Fig. \ref{f:bestcso}. As for the squeezed-state channel, 
the optimization makes the diagram more symmetric around the 
$\eta=1/2$ vertical axis. 
\section{Conclusions}
We analyzed the detrimental effect of loss on narrow-band
quantum-optical channels based on $i)$ heterodyne detection of
coherent states, $ii)$ homodyne detection of squeezed states and
$iii)$ direct detection of number states. We have shown that the
squeezed-state channel and, even more, the number-state channel, are both
easily degraded by loss below the capacity of the coherent-state channel. 
Because of the peculiar form of the master equation for the loss, the
coherent-state channel does not need optimization, and remains as the
most efficient one at sufficiently high power.  
\par The optimization of the squeezed-state channel leads to a sizeable
improvement of the mutual information (over 30\% for $\eta=.15$ at low
power). Correspondingly, the optimal fraction of squeezing photons
rapidly decreases with attenuation. For total average number of photons 
$N < 8$ the squeezed-state channel is always more efficient than the 
coherent-state one, independently on attenuation $\eta$. The
optimization has been performed at constant squeezing, whereas the
problem of optimizing a signal-dependent squeezing is still open.
\par As regards the number-state channel, we applied the Blahut's
recursive algorithm to evaluate the optimal {\em a priori} probability
and the channel capacity. The improvement of the mutual information is
considerable,  achieving $70 \%$ for $\eta=.15$. The optimal {\em a
priori} probability departs from the usual monotonic thermal-like
distribution, and for $\eta\lesssim .6$ it develops gaps of zero
probability at intermediate number of photons. At low power the
optimization of the number-state channel makes its capacity better
than that of the coherent-state channel. 
\par A comprehensive view of the numerical results of this paper is
offered by the optimality capacity diagram in Fig. \ref{f:bestopt}:
there one can find the regions on the loss-power plane where the
coherent-state, the optimized squeezed-state, and the optimized
number-state channels are respectively optimal.  
\section*{Acknowledgments}
We gratefully acknowledge interesting and stimulating discussions with
H. P. Yuen, who also attracted our attention on the main issue of this
paper. 

\newpage

\newpage

\begin{table}[hbt]
\begin{center}
\begin{tabular}{|r|r|r|r|r|r|r|r|r|}
plot & $\eta $ & $N$ & $I_{opt}$ & $I_{th}$ & $I_{coh}$ & \% & 
$\epsilon _P$ & $\epsilon _I$ \\ \hline\hline
a) & .9 & 8.575 & 3.157 & 3.097 & 3.124 & 1.93 & 2$\cdot 10^{-12}$ &  
 1$\cdot 10^{-18}$ \\ \hline
b) & .75 & 2.827 & 1.775 & 1.699 & 1.642 & 4.50 & 4$\cdot 10^{-13}$ &  
 1$\cdot 10^{-18}$ \\ \hline
c) & .6 & 2.414 & 1.340 & 1.218 & 1.292 & 10.03 & 1$\cdot 10^{-8}$ &  
 1$\cdot 10^{-14}$ \\ \hline
d) & .6 & 6.930 & 1.935 & 1.745 & 2.367 & 10.90 & 2$\cdot 10^{-8}$ &  
 6$\cdot 10^{-14}$ \\ \hline
e) & .55 & 2.288 & 1.219 & 1.083 & 1.175 & 12.56 & 8$\cdot 10^{-8}$ &  
 7$\cdot 10^{-13}$ \\ \hline
f) & .55 & 6.729 & 1.803 & 1.595 & 2.233 & 13.07 & 1$\cdot 10^{-7}$ &  
 2$\cdot 10^{-12}$ \\ \hline
g) & .4 & 1.888 & 0.887 & 0.715 & 0.812 & 24.18 & 6$\cdot 10^{-8}$ &  
 2$\cdot 10^{-12}$ \\ \hline
h) & .15 & 4.040 & 0.720 & 0.416 & 0.684 & 73.08 & 8$\cdot 10^{-9}$ &  
 2$\cdot 10^{-13}$ \\ 
\end{tabular}
\end{center}
\caption[fake]{Numerical values relative to the plots a-h of
Fig. \ref{f:pn1}. The table lists the following quantities: 
attenuation factor ($\eta$); average number of photons ($N$); 
mutual information (in bits) ($I_{opt}$) for the optimized number-state 
channel, ($I_{th}$) for the number-state channel with customary thermal 
probability, ($I_{coh}$) for the coherent-state channel; {\em per cent} 
improvement (\%) of the mutual information due to the optimization; 
convergence parameters $\epsilon _P$ and $\epsilon _I$ (see text).}
\label{tab}\end{table}

\end{document}